\documentclass[aps,prl,twocolumn]{revtex4-1}
\pdfoutput=1
\usepackage{hyperref}
\usepackage{graphicx}
\usepackage[applemac]{inputenc}
\usepackage{natbib}
\usepackage{bm}
\usepackage{amssymb}
\usepackage{amsbsy}
\usepackage[usenames]{color}
\usepackage{mathrsfs,mathcomp}
 \usepackage[normalem]{ulem}
\usepackage{amsmath}
\usepackage{amsfonts}
\usepackage{mathrsfs}
\usepackage{subfigure}
\usepackage{ifsym} 
\usepackage{multirow}
\usepackage{textgreek}
\usepackage{textcomp}
\usepackage{gensymb}
\usepackage[english]{babel}

\newcommand{\SIkHz}{\textrm{kHz}}

\newcommand{\SIkg}{\textrm{kg}}

\newcommand{\SIm}{\textrm{m}}

\newcommand{\SImum}{\textrm{\textmu{}m}}

\begin{document}

\title{Three-dimensional phenomena in microbubble acoustic streaming}

\author{Alvaro G. Mar\'in$^1$, Massimiliano Rossi$^1$, Bhargav Rallabandi $^2$, Cheng Wang, Sascha Hilgenfeldt$^2$, Christian J. K\"ahler$^1$} 

\affiliation{$^1$Institut f\"ur Str\"omungsmechanik und Aerodynamik, Bundeswehr University Munich, Germany\\
$^2$ Department of Mechanical Science and Engineering, University of Illinois at Urbana-Champaign, 1206 West Green Street, Urbana, IL 61801, USA.}

\date{\today}

\begin{abstract}
Ultrasound-driven oscillating micro-bubbles have been used as active actuators in microfluidic devices to perform manifold tasks such as mixing, sorting and manipulation of microparticles. A common configuration consists on side-bubbles, created by trapping air pockets in blind channels perpendicular to the main channel direction. 
This configuration consists of acoustically excited bubbles with a semi-cylindrical shape that generate significant streaming flow. Due to the geometry of the channels, such flows have been generally considered as quasi two-dimensional. Similar assumptions are often made in many other microfluidic systems based on \emph{flat} micro-channels. However, in this paper we show that microparticle trajectories actually present a much richer behavior, with particularly strong out-of-plane dynamics in regions close to the microbubble interface. Using Astigmatism Particle Tracking Velocimetry, we reveal that the apparent planar streamlines are actually projections of a streamsurface with a pseudo-toroidal shape. We therefore show that acoustic streaming cannot generally be assumed as a two-dimensional phenomenon in confined systems. The results have crucial consequences for most of the applications involving acoustic streaming as particle trapping, sorting and mixing. 

\end{abstract}

\maketitle

Among the vast amount of microfluidics systems that have been emerging since the late 90's, those involving the use of acoustic fields to enhance flow control (i.e. acoustofluidics) have experienced a great development in recent years due to their practical simplicity, versatility and potentiality for applications as particle separation \cite{Petersson:2005ky, Laurell:2007kg, Augustsson2011}, particle trapping \cite{Hammarstrom:2010iz}, acoustic manipulation of droplets using surface acoustic waves \cite{Friend:2011gf,Franke:2009eg} or acoustic centrifugation \cite{Raghavan:2009fj}. Direct piezoelectric actuation of microchannel walls generates acoustic waves whose interaction with suspended particles through streaming flows or radiation forces has been used to concentrate, trap, or separate them through either acoustic streaming (interaction with the fluid) or radiation forces (interaction with the particles) \cite{Bruus:2011jl,Bruus:2012wp,Muller:2013iy}. However, other type of applications may require higher shear stresses or just a different range of frequencies than those achievable by standing acoustic waves. Instead of actuating directly on the solid boundary of the channel, an alternative is to actuate on the surface of a sessile microbubble that has been deliberately created in the system. The kHz-frequency oscillations of the bubble surface then generate a steady streaming flow of adjustable intensity. Such a technique is particularly effective in soft-wall microfluidic systems such as PDMS (polydimethylsiloxane) devices, where waves in the substrate are strongly damped. Ultrasound-driven microbubble streaming flows has been studied for applications to promote fluid transport \cite{Marmottant:2006hg}, manipulation and poration of vesicles \cite{Marmottant:2008fg}, and more recently for applications such as particle separation and trapping \cite{Wang:2011ip,Wang:2012hg}.

\begin{figure} [h!]	
\includegraphics[width=\columnwidth,resolution=100]{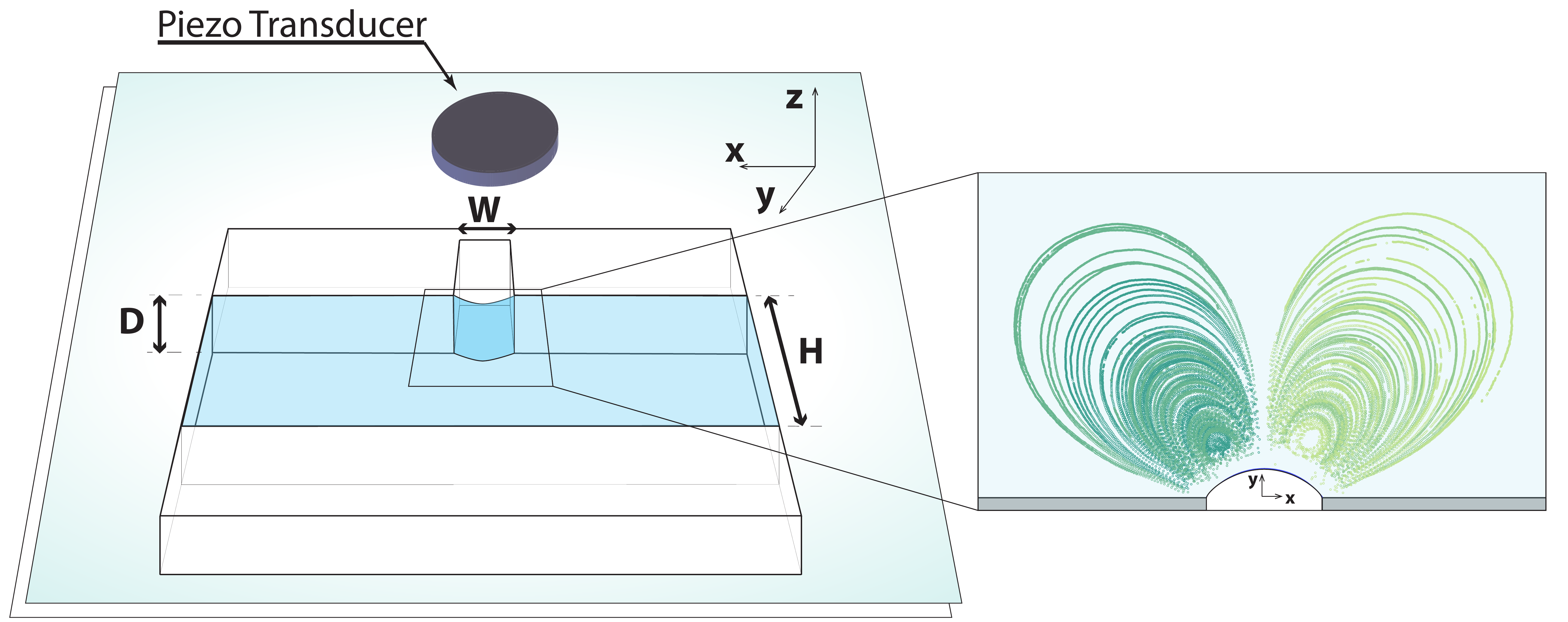}
\caption{Typical experimental set-up. Left:  The PDMS channel and the piezoelectric actuator are bonded to a glass slide. Right: Top View of the semi-cylindrical bubble in the microchannel and visualization of the acoustic streaming flow through different experimental particle trajectories.}\label{figsetup}
\end{figure}

One of the simplest and most natural configurations to insert stable sessile microbubbles in PDMS microchannels is to create blind side channels. When the channel is filled with water, gas is retained in the pocket which then adopts a semicylindrical shape (Figure \ref{figsetup}). Due to the lithography-based micro fabrication, a 2D planar geometry is imposed and a reasonable assumption is that the two-dimensional shape of the channel and its semi-cylindrical bubble shape damps the third component of the flow field, leaving a 2D flow profile. The assumption has been proven valid enough to study general aspects of the phenomenon as the typical frequency resonances \cite{Wang:2013cy} or the two-dimensional streaming flow by using asymptotic models \cite{Rallabandi:2013eb}.

\begin{figure*}[t!]
\includegraphics[width=1.8\columnwidth]{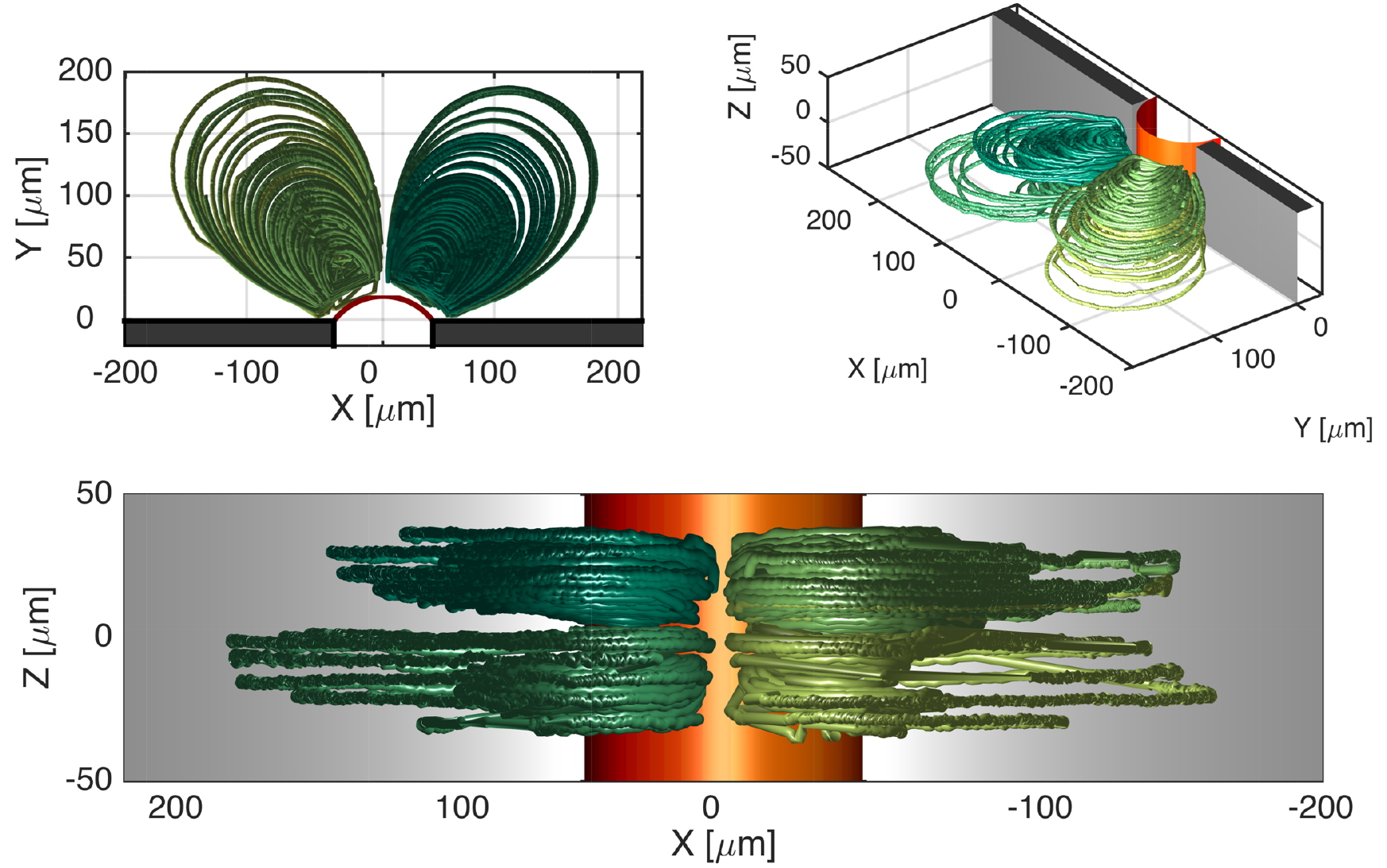}
\caption{Trajectories of four different particles. Top left: Projection of particles trajectories in the XY plane. Top right: Orthographic projection of the particle trajectories. Bottom: Projection of particles trajectories in the XZ plane}\label{fig2}
\end{figure*}

In this paper, we demonstrate through experimental measurements that particles do experience off-plane dynamics in the vicinity of the micro-bubble, which reveals an unexpected degree of complexity in confined fluidic systems driven by acoustic streaming. In the next lines we will describe the dynamics of the complex flow, point out the consequences for applications, as well as determine in which circumstances one can ignore the three-dimensional features of the flow. In order to inquire on the nature of the flow, a three-dimensional solution of the Stokes equation showing the same trend can be constructed, which compares well with the experimental results. A more detailed description of the theoretical model will be shown in a companion paper \cite{rallabandi2014}.

\paragraph{Experimental Setup.- } The microfluidic device is made of Polydimethilsiloxane (PDMS) by soft lithography: The PMDS components are blended at a 10:1 ratio, degassed, and poured into a SU-8 mold to cross-link for 24 hours. The cured PMDS is  then bonded to a glass slide via oxygen plasma treatment. Typical microchannels (see Figure \ref{figsetup}) have a height D=100 \SImum, and a width H=1000 \SImum, while the blind channel holding the semi-cylindrical microbubble has a typical width W=80 \SImum. The microbubble is driven by ultrasound actuation through a piezoelectric transducer (10 mm diameter, Physik Instrumente, Germany) glued to the glass slide, next to the PDMS channel (Figure \ref{figsetup}). An amplified sinusoidal signal is sent to the transducer through a function generator GW-Instek AFG-2125. All experiments shown in the paper have been made at a typical frequency of $\sim 20$ \SIkHz, appropriate for strong excitation of microbubble of the present size \cite{Wang:2013cy}. In order to acquire more data at sufficient resolution, the flow can be slowed down by choosing a lower voltage at the piezo transducer.

\begin{figure*}[t!]
\includegraphics[width=2\columnwidth]{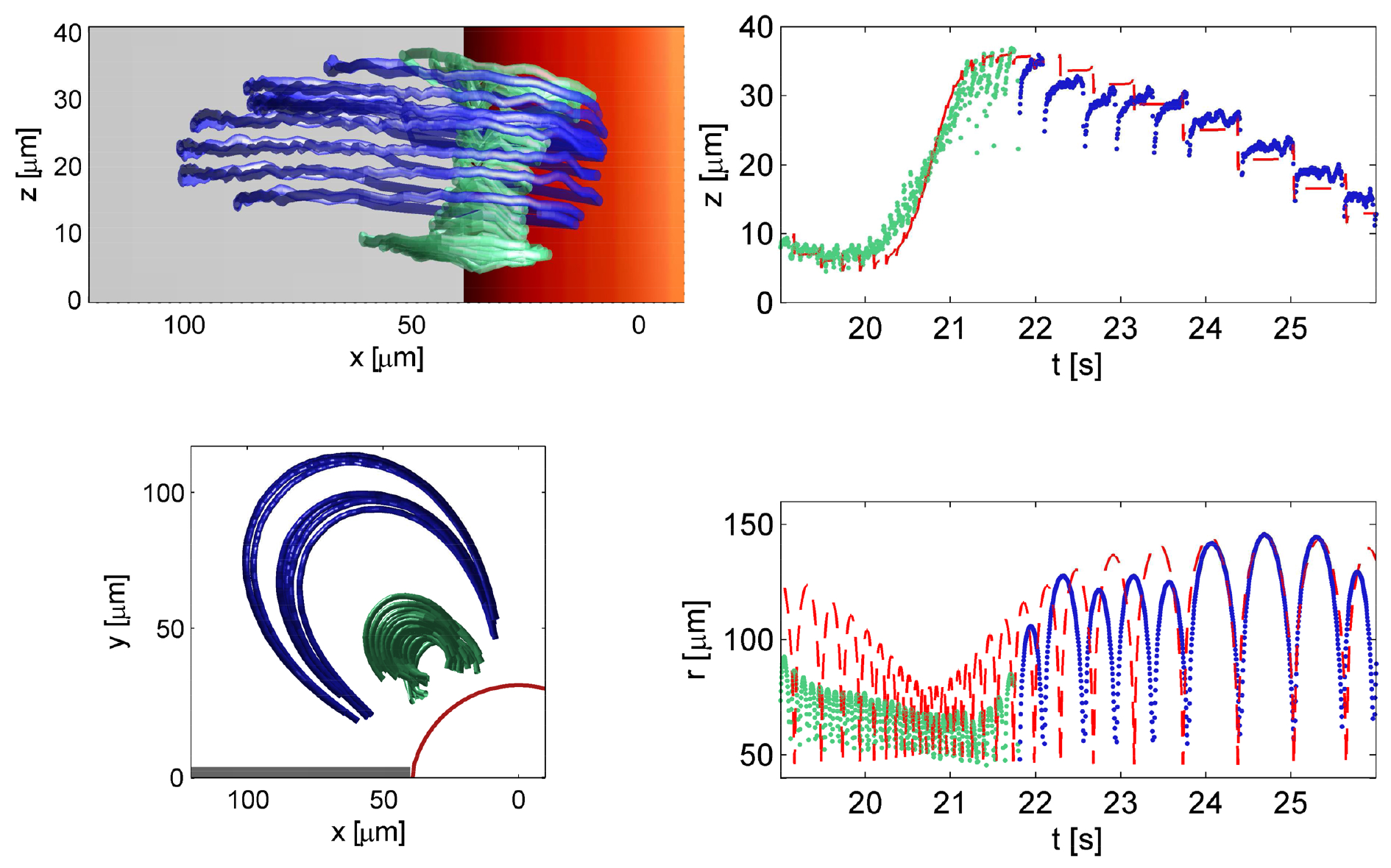}
\caption{Detail of a typical particle trajectory. The green and blue paths depict respectively the \emph{inner} and \emph{outer} particle loops, in which the particle travels respectively \emph{away} from and \emph{towards} the mid plane $z=0$. Top left: Projection of the particle in the ZX plane and in the XY plane (bottom left). Top right: Particle vertical distance from the system's midplane $z=0$, notice the jumps that the particle performs in the closest position to the bubble's surface. The red discontinuous line shows the results from simulations. Bottom right: Radial distance $r=\sqrt{x^2+y^2}$ of the particle to the bubble's symmetry in time}\label{fig3}
\end{figure*}

\paragraph{3D particle tracking.-} The particle trajectories and velocities are measured using astigmatism particle tracking velocimetry (APTV) ~\cite{Cierpka2011, rossi2014optimization}. APTV is a single-camera particle-tracking method in which an astigmatic aberration is introduced in the optical system by means of a cylindrical lens placed in front of the camera sensor. Consequently, an image of a spherical particle obtained in such a system shows a characteristic elliptical shape unequivocally related to its depth-position $z$. The images of the particles in the microfluidic chip are taken using an inverted microscope Zeiss Axiovert in combination with a high-speed CCD camera PCO.dimax at recording speeds in the range from 500 to 1000 fps. The optical arrangement consists of a Zeiss Plan-Neofluar 20x/0.4 microscope objective lens and a cylindrical lens with focal length $f_\mathrm{cyl} = 150$~mm placed in front of the CCD sensor of the camera. Monodisperse fluorescent spherical polystyrene particles with nominal diameters of $2~\SImum$ (Microparticles GmbH) were used for the experiments ($\rho^{{}}_\mathrm{ps} = 1050~\SIkg~\SIm^{-3}$). Illumination is provided by a continuous diode-pumped laser with 2~W power at 532~nm wavelength. This configuration provides a measurement volume of $600\times 600\times 120~\SImum^3$ with an estimated uncertainty in the particle position of $\pm 1~\SImum$ in the $z$-direction and less than $\pm 0.1~\SImum$ in the $x$- and $y$-direction. The particle trajectories detected for this frequency vary in length and in time scale, and add up to more than 50 different trajectories. More details about the experimental configuration and uncertainty estimation of the APTV system can be found in reference \cite{rossi2014optimization}.

\paragraph{Results.-} Figure \ref{fig2} shows measured trajectories corresponding to four selected particles. Each trajectory occupies different volumes separated by two planes of symmetry: the mid-plane $z=0$ and the bubble's sagittal plane $y=0$. A third symmetry plane would be expected if the bubble were perfectly cylindrical. Most of the observed particles are confined in between the symmetry planes, but particles crossing the symmetry planes are also eventually observed.

Figure \ref{fig3} illustrates a portion of a trajectory: particles follow an almost flat trajectory at large $r$ but experience discrete ``jumps'' into a different plane every time the particle's orbit reaches its perigee, i.e. when the particle comes to its closest distance to the bubble's surface ($r_{min}$).  Particle trajectories typically follow a periodic pattern similar to the one depicted in Figure \ref{fig3}: The particle starts close to the mid-plane $z=0$ with series of fast loops characterized by short periods, short apogees (farthest position from the bubble's surface $r_{max}$) and large perigees (trajectory in green in Figure \ref{fig3}). When passing through the perigee, the particle suffers small jumps \emph{away} from the mid-plane. As the particle approaches the channel ceiling/floor, the orbit reduces the perigee, enlarges its apogee and consequently its period becomes longer (blue trajectory in Figure \ref{fig3}). Now, the particle jumps \emph{towards} the mid-plane every time the it hits its perigee. As the particle approaches the mid-plane again, the orbit decreases its apogee, reduces its perigee, and will repeat the same pattern again. Since particles suffer strong velocity gradients, when choosing the experimental settings one faces a dilemma: either to resolve the short time-scale particle dynamics or the long time-scale dynamics. Since our objective is to get a general view of the particle orbit, we chose the latter option. Therefore, note from the bottom plots in Figure \ref{fig3} that the selected experimental settings do not allow us to resolve experimentally the portion of the particle trajectory closest to the bubble and therefore the streamlines are not closed. 

\begin{figure}[h]
\includegraphics[width=\columnwidth]{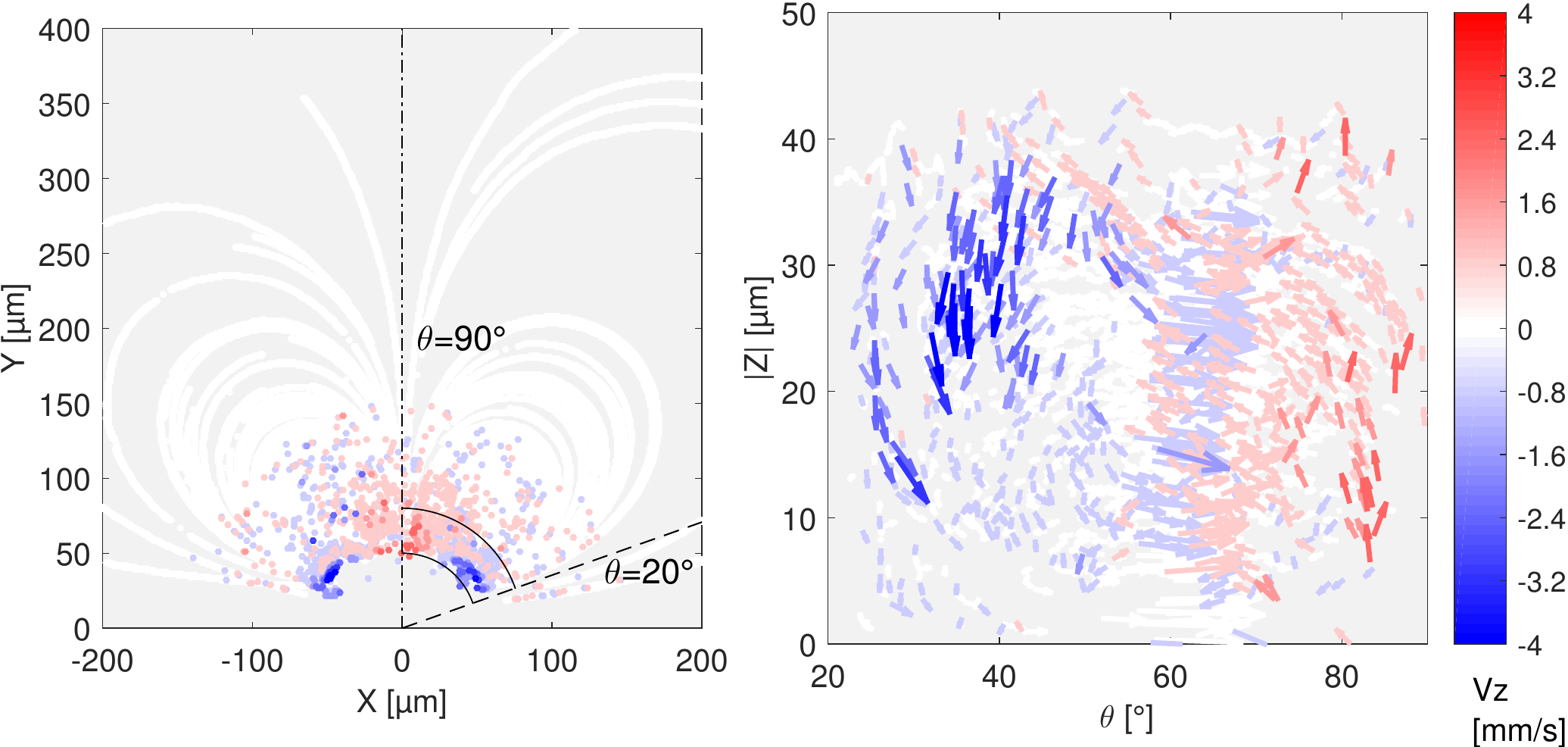}
\includegraphics[width=\columnwidth]{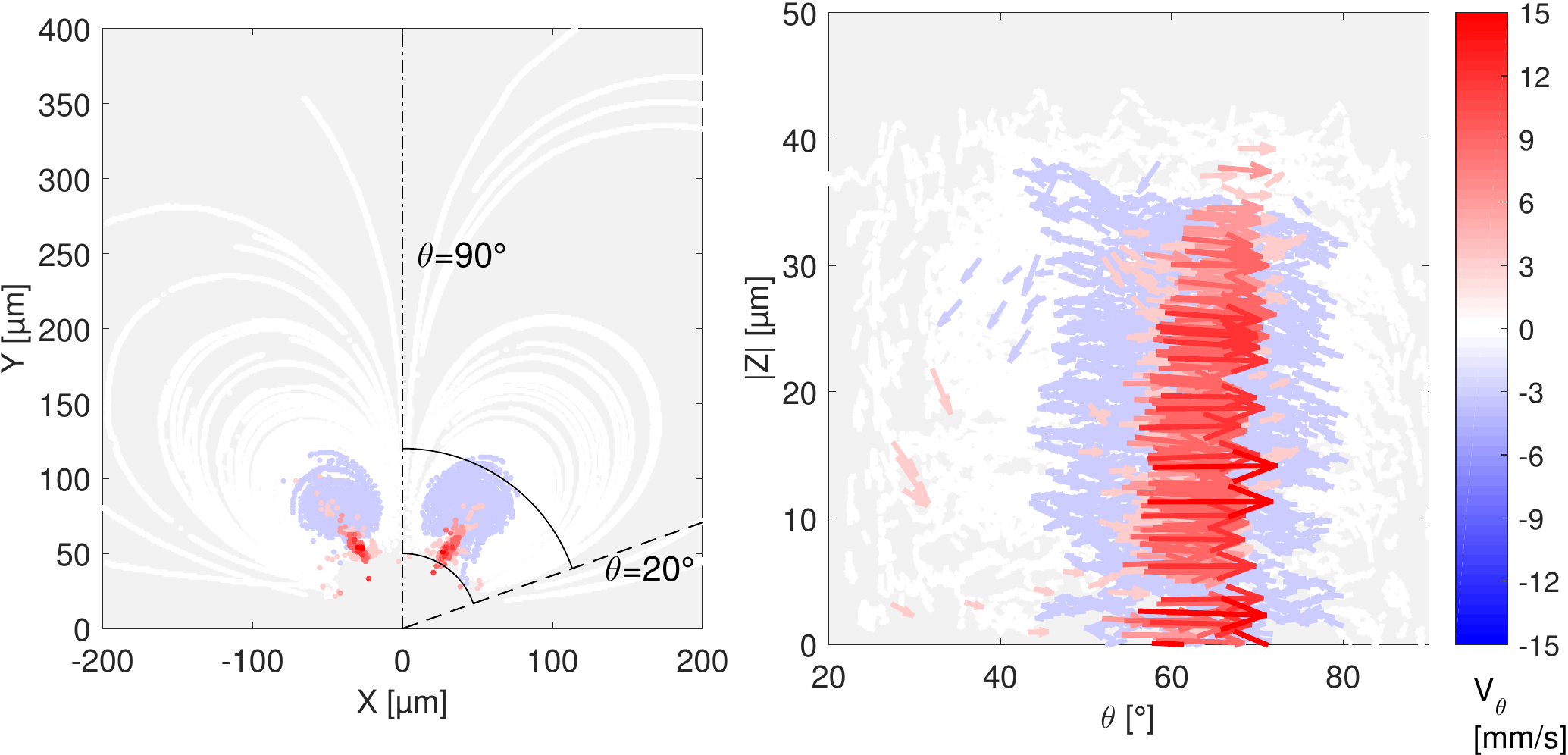}
\includegraphics[width=\columnwidth]{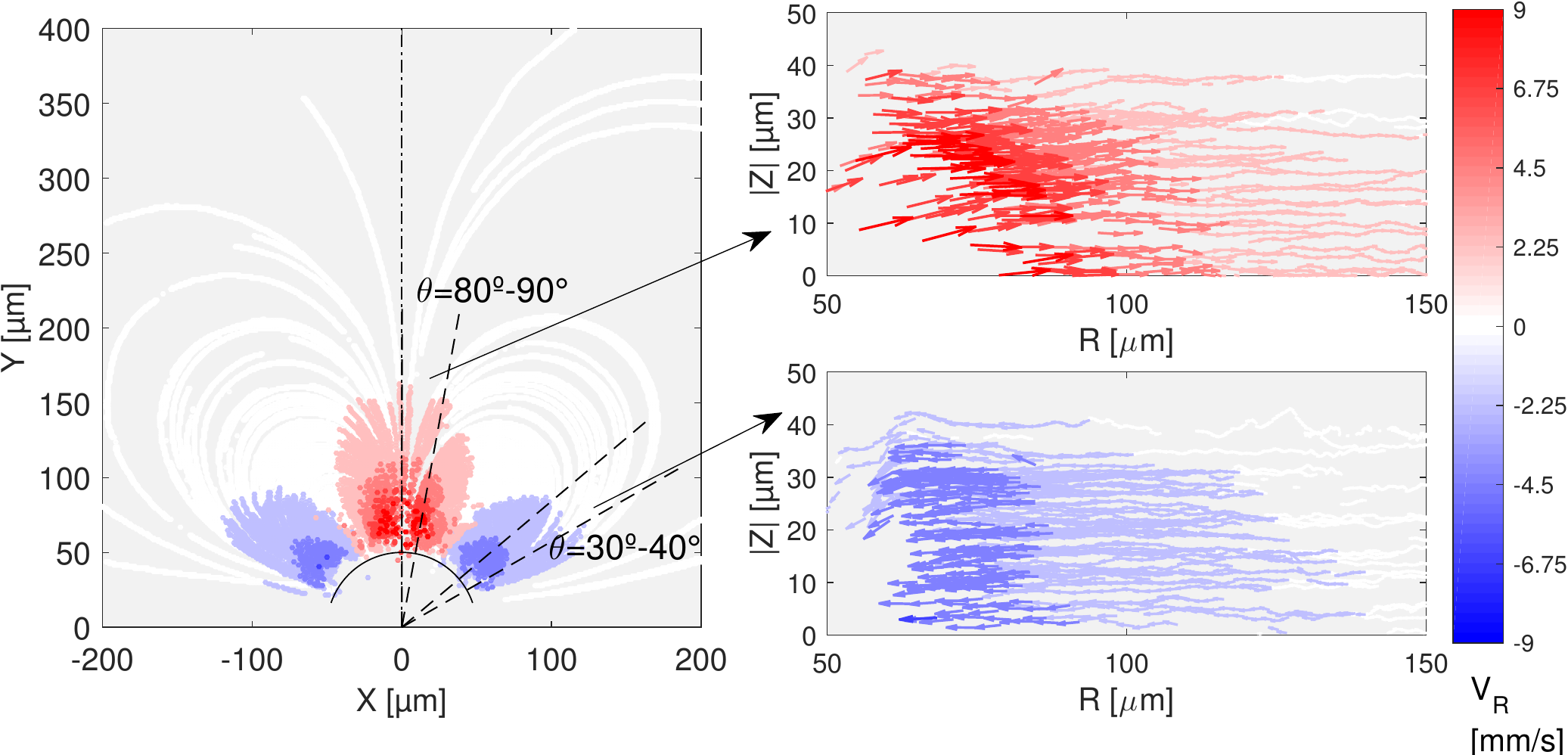}

\caption{Particle Velocity distributions. Top: off plane velocity component $V_z$, note that they only become significant in the vicinity of the bubble's surface. The arcs drawn in the top left Figure represents the section plotted in the top right plot. Middle: Azimuthal velocity component $V_{\theta}$. Bottom: Radial velocity $V_r$ with maximum values at the sagittal plane and at approximately $\theta=30 \degree$ (bottom right plots). }\label{fig4}
\end{figure}

Another remarkable symmetry becomes visible when observing the velocity field in detail in Figure \ref{fig4}: as particles approach the bubble at low angles $\theta$, they are advected into the bubble's horizontal plane $z=0$ at $\theta<60\degree$ and immediately after at $\theta>60\degree$ away from it. Once the particle leaves the bubble behind, there is a net displacement in its $z-$position whose direction and value depends strongly on the relative position to the bubble. Interestingly enough, the critical angle $\theta\approx60\degree$ at which the value of $V_z$ changes sign corresponds also to the angle at which $V_r$ changes sign, and to the position at which $V_{\theta}$ reaches its maximum value (Figure \ref{fig4}).  It is interesting to note that in the horizontal plane $z=0$, $V_{\theta}$ and $V_r$ reach their maximum values, while $V_z$ reaches a minimum.
In the data shown, the maximum Reynolds number that a 2-micron-particle experiences is $Re_p\sim 10^{-2}$, and therefore inertial effects do not play any role. However, even though the flow is clearly viscous, stresses can be significant. For example, the maximum shear rates experienced at the top/bottom walls can be as high as 500 1/s, while in the bulk they reach 1000 1/s. Note that mechanisms as shear-induced migration cannot be invoked to explain the particle dynamics due to the low particle seeding employed in our experiments.
It was normally assumed that the streamlines were two-dimensional and mostly closed \cite{Versluis:2006vq}. However when observing the planar projection, one can observe particle paths crossing streamlines. It was normally concluded that the particle was perturbed by the bubble surface oscillations, forcing the particle to change streamlines \cite{Wang:2012hg}. Although such a mechanism is valid and probably occurs as well, the present results show that apparent streamline crossings are also caused by the particle following a complex 3D \emph{streamsurface} that has a toroidal shape (Figure \ref{fig3}). That explains why apparent streamline crossing occurs even for very small tracer particles. Interestingly enough, particles are very rarely seen making a full and closed 3D trajectory. The most likely reason is that the time required to go through the full three-dimensional orbit (a full period) is much longer than our observation time. Still, streamline crossing cannot be totally neglected since the chances of a perturbation to occur while the particle travels are higher for longer periods. 

\paragraph{Discussion.-} 
Secondary flows unavoidably appear in three-dimensional systems even at low Reynolds numbers when a streamwise component of vorticity is developed in the presence of a boundary layer. A classical example of secondary flow is the so-called tea leaf effect \cite{einstein1926ursache}, which explains why particles concentrate in the bottom-center of a stirred tea cup. But they are also present when a pipe flow enters a bend or over an immersed rotating disc. In our particular case, in addition to this common phenomenon we must add the effect of the oscillations along the axis of the cylindrical bubble, which cannot be neglected. These effects can be taken into account by a modification of the two-dimensional bubble streaming solution, which is governed by a stream function $\psi_{2d}(r,\theta)$ \cite{Rallabandi:2013eb,rallabandi2014}. Due to the small Reynolds number associated with the streaming, it is appropriate to seek a linear superposition involving $\psi_{2d}$ and axial solutions of the Stokes equations satisfying no penetration conditions at the bubble and the walls of the microchannel, as well as the appropriate symmetries about $x=0$ and $z=0$ observed in the experiment. The lowest order axial solution (weakest radial decay) is axisymmetric and is governed by a Stokes streamfunction $\psi_{ax}$ of the form
\begin{equation}
 \psi_{ax}(r,z) = u_s a\, \frac{r}{a}\left(\frac{\mathrm{K}_2(2 \pi r/D)}{\mathrm{K}_2(2 \pi a/D)} -  \frac{\mathrm{K}_0(2 \pi r/D)}{\mathrm{K}_0(2 \pi a/D)}\right) \sin (2\pi z/D).
\end{equation}
Here, $u_s$ is the characteristic scale of the streaming near the bubble surface (quadratic in driving voltage), $a=W/2$ is the bubble radius, and $\mathrm{K}_{\nu}$ represents a modified Bessel function of the second kind.  The velocity field due to the superposition of 2D and axisymmetric modes is formally given by
\begin{equation}\label{superposition}
\mathbf{u}(r,\theta,z) = \nabla \times \left\{\psi_{2d}(r,\theta)\, \mathbf{\hat{z}} + c_{ax}\psi_{ax}(r,z)\, \mathbf{\hat{\theta}} \right\}
\end{equation}
where $0<c_{ax}\ll 1$ sets the relative strength of the axial flow and is used as a fit parameter between the theory and experiment. Figure \ref{fig3} shows that fluid trajectories under the linear superposition \eqref{superposition} qualitatively agree with experimentally measured particle trajectories over short ($\sim$ ms) and long ($\sim$ s) time scales.

The results here shown may also have crucial consequences in stirring/mixing in microfluidic viscous flows. From Aref \cite{aref1984stirring} and Ottino \cite{ottino1989kinematics} we learned that, in order to enhance mixing in viscous flows, one must generate striations or tendrils thin enough to increase the value of the gradients and to allow diffusion to transfer mass in reasonable time scales. In this sense, 3D flows can achieve better mixing rates even in a passive way since they stretch and fold the liquid more efficiently, as has been shown by the Herringbone mixer \cite{Stroock:2002vd} or the 3D serpentine \cite{Liu:2000iu}. The three-dimensional character of streaming flow from semi-cylindrical bubble contributes to the good mixing that can be already achieved in the liquid surrounding one single bubble. Mixing can be further enhanced by simply adjusting the bubble's aspect ratio to enhance 3D effects, in addition to modulating its actuating frequency or adding more bubbles in different 3D positions \cite{Wang:2013cy}.

\paragraph{Conclusions and Outlook.-}

In this paper we have shown that the previously assumed 2D character of acoustic streaming by semi-cylindrical bubbles does not hold when particles approach the bubble's vicinity. Instead, 3D flow effects are present that have crucial consequences on the particle trajectories and therefore on the fluid flow surrounding the bubble. The mechanisms driving such a flow must necessarily be connected with the presence of axial bubble oscillations and their necessary fulfillment of the boundary conditions. 


\paragraph{Acknowledgments.-}
Abundant discussions with Rune Barnkob and Andreas Volk are greatly acknowledged. The authors acknowledge financial support by the Deutsche Forschungsgemeinschaft grant KA 1808/12-1 and KA 1808/13-1.


\begin{thebibliography}{25}%
\makeatletter
\providecommand \@ifxundefined [1]{%
 \@ifx{#1\undefined}
}%
\providecommand \@ifnum [1]{%
 \ifnum #1\expandafter \@firstoftwo
 \else \expandafter \@secondoftwo
 \fi
}%
\providecommand \@ifx [1]{%
 \ifx #1\expandafter \@firstoftwo
 \else \expandafter \@secondoftwo
 \fi
}%
\providecommand \natexlab [1]{#1}%
\providecommand \enquote  [1]{``#1''}%
\providecommand \bibnamefont  [1]{#1}%
\providecommand \bibfnamefont [1]{#1}%
\providecommand \citenamefont [1]{#1}%
\providecommand \href@noop [0]{\@secondoftwo}%
\providecommand \href [0]{\begingroup \@sanitize@url \@href}%
\providecommand \@href[1]{\@@startlink{#1}\@@href}%
\providecommand \@@href[1]{\endgroup#1\@@endlink}%
\providecommand \@sanitize@url [0]{\catcode `\\12\catcode `\$12\catcode
  `\&12\catcode `\#12\catcode `\^12\catcode `\_12\catcode `\%12\relax}%
\providecommand \@@startlink[1]{}%
\providecommand \@@endlink[0]{}%
\providecommand \url  [0]{\begingroup\@sanitize@url \@url }%
\providecommand \@url [1]{\endgroup\@href {#1}{\urlprefix }}%
\providecommand \urlprefix  [0]{URL }%
\providecommand \Eprint [0]{\href }%
\providecommand \doibase [0]{http://dx.doi.org/}%
\providecommand \selectlanguage [0]{\@gobble}%
\providecommand \bibinfo  [0]{\@secondoftwo}%
\providecommand \bibfield  [0]{\@secondoftwo}%
\providecommand \translation [1]{[#1]}%
\providecommand \BibitemOpen [0]{}%
\providecommand \bibitemStop [0]{}%
\providecommand \bibitemNoStop [0]{.\EOS\space}%
\providecommand \EOS [0]{\spacefactor3000\relax}%
\providecommand \BibitemShut  [1]{\csname bibitem#1\endcsname}%
\let\auto@bib@innerbib\@empty
\bibitem [{\citenamefont {Petersson}\ \emph {et~al.}(2005)\citenamefont
  {Petersson}, \citenamefont {Nilsson}, \citenamefont {J{\"o}nsson},\ and\
  \citenamefont {Laurell}}]{Petersson:2005ky}%
  \BibitemOpen
  \bibfield  {author} {\bibinfo {author} {\bibfnamefont {F.}~\bibnamefont
  {Petersson}}, \bibinfo {author} {\bibfnamefont {A.}~\bibnamefont {Nilsson}},
  \bibinfo {author} {\bibfnamefont {H.}~\bibnamefont {J{\"o}nsson}}, \ and\
  \bibinfo {author} {\bibfnamefont {T.}~\bibnamefont {Laurell}},\ }\href@noop
  {} {\bibfield  {journal} {\bibinfo  {journal} {Anal. Chem.}\ }\textbf
  {\bibinfo {volume} {77}},\ \bibinfo {pages} {1216} (\bibinfo {year}
  {2005})}\BibitemShut {NoStop}%
\bibitem [{\citenamefont {Laurell}\ \emph {et~al.}(2007)\citenamefont
  {Laurell}, \citenamefont {Petersson},\ and\ \citenamefont
  {Nilsson}}]{Laurell:2007kg}%
  \BibitemOpen
  \bibfield  {author} {\bibinfo {author} {\bibfnamefont {T.}~\bibnamefont
  {Laurell}}, \bibinfo {author} {\bibfnamefont {F.}~\bibnamefont {Petersson}},
  \ and\ \bibinfo {author} {\bibfnamefont {A.}~\bibnamefont {Nilsson}},\ }\href
  {\doibase 10.1039/b601326k} {\bibfield  {journal} {\bibinfo  {journal} {Chem.
  Soc. Rev.}\ }\textbf {\bibinfo {volume} {36}},\ \bibinfo {pages} {492}
  (\bibinfo {year} {2007})}\BibitemShut {NoStop}%
\bibitem [{\citenamefont {Augustsson}\ \emph {et~al.}(2011)\citenamefont
  {Augustsson}, \citenamefont {Barnkob}, \citenamefont {Wereley}, \citenamefont
  {Bruus},\ and\ \citenamefont {Laurell}}]{Augustsson2011}%
  \BibitemOpen
  \bibfield  {author} {\bibinfo {author} {\bibfnamefont {P.}~\bibnamefont
  {Augustsson}}, \bibinfo {author} {\bibfnamefont {R.}~\bibnamefont {Barnkob}},
  \bibinfo {author} {\bibfnamefont {S.~T.}\ \bibnamefont {Wereley}}, \bibinfo
  {author} {\bibfnamefont {H.}~\bibnamefont {Bruus}}, \ and\ \bibinfo {author}
  {\bibfnamefont {T.}~\bibnamefont {Laurell}},\ }\href {\doibase
  10.1039/c1lc20637k} {\bibfield  {journal} {\bibinfo  {journal} {Lab Chip}\
  }\textbf {\bibinfo {volume} {11}},\ \bibinfo {pages} {4152} (\bibinfo {year}
  {2011})}\BibitemShut {NoStop}%
\bibitem [{\citenamefont {Hammarstr{\"o}m}\ \emph {et~al.}(2010)\citenamefont
  {Hammarstr{\"o}m}, \citenamefont {Evander}, \citenamefont {Barbeau},
  \citenamefont {Bruzelius}, \citenamefont {Larsson}, \citenamefont {Laurell},\
  and\ \citenamefont {Nilsson}}]{Hammarstrom:2010iz}%
  \BibitemOpen
  \bibfield  {author} {\bibinfo {author} {\bibfnamefont {B.}~\bibnamefont
  {Hammarstr{\"o}m}}, \bibinfo {author} {\bibfnamefont {M.}~\bibnamefont
  {Evander}}, \bibinfo {author} {\bibfnamefont {H.}~\bibnamefont {Barbeau}},
  \bibinfo {author} {\bibfnamefont {M.}~\bibnamefont {Bruzelius}}, \bibinfo
  {author} {\bibfnamefont {J.}~\bibnamefont {Larsson}}, \bibinfo {author}
  {\bibfnamefont {T.}~\bibnamefont {Laurell}}, \ and\ \bibinfo {author}
  {\bibfnamefont {J.}~\bibnamefont {Nilsson}},\ }\href {\doibase
  10.1039/c004504g} {\bibfield  {journal} {\bibinfo  {journal} {Lab Chip}\
  }\textbf {\bibinfo {volume} {10}},\ \bibinfo {pages} {2251} (\bibinfo {year}
  {2010})}\BibitemShut {NoStop}%
\bibitem [{\citenamefont {Friend}\ and\ \citenamefont
  {Yeo}(2011)}]{Friend:2011gf}%
  \BibitemOpen
  \bibfield  {author} {\bibinfo {author} {\bibfnamefont {J.}~\bibnamefont
  {Friend}}\ and\ \bibinfo {author} {\bibfnamefont {L.}~\bibnamefont {Yeo}},\
  }\href {\doibase 10.1103/RevModPhys.83.647} {\bibfield  {journal} {\bibinfo
  {journal} {Rev. Mod. Phys.}\ }\textbf {\bibinfo {volume} {83}},\ \bibinfo
  {pages} {647} (\bibinfo {year} {2011})}\BibitemShut {NoStop}%
\bibitem [{\citenamefont {Franke}\ \emph {et~al.}(2009)\citenamefont {Franke},
  \citenamefont {Abate}, \citenamefont {Weitz},\ and\ \citenamefont
  {Wixforth}}]{Franke:2009eg}%
  \BibitemOpen
  \bibfield  {author} {\bibinfo {author} {\bibfnamefont {T.}~\bibnamefont
  {Franke}}, \bibinfo {author} {\bibfnamefont {A.~R.}\ \bibnamefont {Abate}},
  \bibinfo {author} {\bibfnamefont {D.~A.}\ \bibnamefont {Weitz}}, \ and\
  \bibinfo {author} {\bibfnamefont {A.}~\bibnamefont {Wixforth}},\ }\href
  {\doibase 10.1039/b906819h} {\bibfield  {journal} {\bibinfo  {journal} {Lab
  Chip}\ }\textbf {\bibinfo {volume} {9}},\ \bibinfo {pages} {2625} (\bibinfo
  {year} {2009})}\BibitemShut {NoStop}%
\bibitem [{\citenamefont {Raghavan}\ \emph {et~al.}(2009)\citenamefont
  {Raghavan}, \citenamefont {Friend},\ and\ \citenamefont
  {Yeo}}]{Raghavan:2009fj}%
  \BibitemOpen
  \bibfield  {author} {\bibinfo {author} {\bibfnamefont {R.~V.}\ \bibnamefont
  {Raghavan}}, \bibinfo {author} {\bibfnamefont {J.~R.}\ \bibnamefont
  {Friend}}, \ and\ \bibinfo {author} {\bibfnamefont {L.~Y.}\ \bibnamefont
  {Yeo}},\ }\href {\doibase 10.1007/s10404-009-0452-3} {\bibfield  {journal}
  {\bibinfo  {journal} {Microfluid Nanofluid}\ }\textbf {\bibinfo {volume}
  {8}},\ \bibinfo {pages} {73} (\bibinfo {year} {2009})}\BibitemShut {NoStop}%
\bibitem [{\citenamefont {Bruus}(2011)}]{Bruus:2011jl}%
  \BibitemOpen
  \bibfield  {author} {\bibinfo {author} {\bibfnamefont {H.}~\bibnamefont
  {Bruus}},\ }\href {\doibase 10.1039/c1lc20770a} {\bibfield  {journal}
  {\bibinfo  {journal} {Lab Chip}\ }\textbf {\bibinfo {volume} {12}},\ \bibinfo
  {pages} {20} (\bibinfo {year} {2011})}\BibitemShut {NoStop}%
\bibitem [{\citenamefont {Bruus}(2012)}]{Bruus:2012wp}%
  \BibitemOpen
  \bibfield  {author} {\bibinfo {author} {\bibfnamefont {H.}~\bibnamefont
  {Bruus}},\ }\href
  {http://pubs.rsc.org/en/content/articlehtml/2012/lc/c2lc21068a} {\bibfield
  {journal} {\bibinfo  {journal} {Lab Chip}\ }\textbf {\bibinfo {volume} {12}}
  (\bibinfo {year} {2012})}\BibitemShut {NoStop}%
\bibitem [{\citenamefont {Muller}\ \emph {et~al.}(2013)\citenamefont {Muller},
  \citenamefont {Rossi}, \citenamefont {Marin}, \citenamefont {Barnkob},
  \citenamefont {Augustsson}, \citenamefont {Laurell}, \citenamefont
  {K{\"a}hler},\ and\ \citenamefont {Bruus}}]{Muller:2013iy}%
  \BibitemOpen
  \bibfield  {author} {\bibinfo {author} {\bibfnamefont {P.~B.}\ \bibnamefont
  {Muller}}, \bibinfo {author} {\bibfnamefont {M.}~\bibnamefont {Rossi}},
  \bibinfo {author} {\bibfnamefont {A.~G.}\ \bibnamefont {Marin}}, \bibinfo
  {author} {\bibfnamefont {R.}~\bibnamefont {Barnkob}}, \bibinfo {author}
  {\bibfnamefont {P.}~\bibnamefont {Augustsson}}, \bibinfo {author}
  {\bibfnamefont {T.}~\bibnamefont {Laurell}}, \bibinfo {author} {\bibfnamefont
  {C.~J.}\ \bibnamefont {K{\"a}hler}}, \ and\ \bibinfo {author} {\bibfnamefont
  {H.}~\bibnamefont {Bruus}},\ }\href {\doibase 10.1103/PhysRevE.88.023006}
  {\bibfield  {journal} {\bibinfo  {journal} {Physical Review E}\ }\textbf
  {\bibinfo {volume} {88}},\ \bibinfo {pages} {023006} (\bibinfo {year}
  {2013})}\BibitemShut {NoStop}%
\bibitem [{\citenamefont {Marmottant}\ \emph
  {et~al.}(2006{\natexlab{a}})\citenamefont {Marmottant}, \citenamefont
  {Raven}, \citenamefont {Gardeniers}, \citenamefont {Bomer},\ and\
  \citenamefont {Hilgenfeldt}}]{Marmottant:2006hg}%
  \BibitemOpen
  \bibfield  {author} {\bibinfo {author} {\bibfnamefont {P.}~\bibnamefont
  {Marmottant}}, \bibinfo {author} {\bibfnamefont {J.-P.}\ \bibnamefont
  {Raven}}, \bibinfo {author} {\bibfnamefont {H.~J. G.~E.}\ \bibnamefont
  {Gardeniers}}, \bibinfo {author} {\bibfnamefont {J.~G.}\ \bibnamefont
  {Bomer}}, \ and\ \bibinfo {author} {\bibfnamefont {S.}~\bibnamefont
  {Hilgenfeldt}},\ }\href {\doibase 10.1017/S0022112006002746} {\bibfield
  {journal} {\bibinfo  {journal} {J. Fluid Mech.}\ }\textbf {\bibinfo {volume}
  {568}},\ \bibinfo {pages} {109} (\bibinfo {year}
  {2006}{\natexlab{a}})}\BibitemShut {NoStop}%
\bibitem [{\citenamefont {Marmottant}\ \emph {et~al.}(2008)\citenamefont
  {Marmottant}, \citenamefont {Biben},\ and\ \citenamefont
  {Hilgenfeldt}}]{Marmottant:2008fg}%
  \BibitemOpen
  \bibfield  {author} {\bibinfo {author} {\bibfnamefont {P.}~\bibnamefont
  {Marmottant}}, \bibinfo {author} {\bibfnamefont {T.}~\bibnamefont {Biben}}, \
  and\ \bibinfo {author} {\bibfnamefont {S.}~\bibnamefont {Hilgenfeldt}},\
  }\href {\doibase 10.1098/rspa.2007.0362} {\bibfield  {journal} {\bibinfo
  {journal} {Proceedings of the Royal Society A: Mathematical, Physical and
  Engineering Sciences}\ }\textbf {\bibinfo {volume} {464}},\ \bibinfo {pages}
  {1781} (\bibinfo {year} {2008})}\BibitemShut {NoStop}%
\bibitem [{\citenamefont {Wang}\ \emph {et~al.}(2011)\citenamefont {Wang},
  \citenamefont {Jalikop},\ and\ \citenamefont {Hilgenfeldt}}]{Wang:2011ip}%
  \BibitemOpen
  \bibfield  {author} {\bibinfo {author} {\bibfnamefont {C.}~\bibnamefont
  {Wang}}, \bibinfo {author} {\bibfnamefont {S.~V.}\ \bibnamefont {Jalikop}}, \
  and\ \bibinfo {author} {\bibfnamefont {S.}~\bibnamefont {Hilgenfeldt}},\
  }\href {\doibase 10.1063/1.3610940.6} {\bibfield  {journal} {\bibinfo
  {journal} {Appl. Phys. Lett.}\ }\textbf {\bibinfo {volume} {99}},\ \bibinfo
  {pages} {034101} (\bibinfo {year} {2011})}\BibitemShut {NoStop}%
\bibitem [{\citenamefont {Wang}\ \emph {et~al.}(2012)\citenamefont {Wang},
  \citenamefont {Jalikop},\ and\ \citenamefont {Hilgenfeldt}}]{Wang:2012hg}%
  \BibitemOpen
  \bibfield  {author} {\bibinfo {author} {\bibfnamefont {C.}~\bibnamefont
  {Wang}}, \bibinfo {author} {\bibfnamefont {S.~V.}\ \bibnamefont {Jalikop}}, \
  and\ \bibinfo {author} {\bibfnamefont {S.}~\bibnamefont {Hilgenfeldt}},\
  }\href {\doibase 10.1063/1.3654949} {\bibfield  {journal} {\bibinfo
  {journal} {Biomicrofluidics}\ }\textbf {\bibinfo {volume} {6}},\ \bibinfo
  {pages} {012801} (\bibinfo {year} {2012})}\BibitemShut {NoStop}%
\bibitem [{\citenamefont {Wang}\ \emph {et~al.}(2013)\citenamefont {Wang},
  \citenamefont {Rallabandi},\ and\ \citenamefont {Hilgenfeldt}}]{Wang:2013cy}%
  \BibitemOpen
  \bibfield  {author} {\bibinfo {author} {\bibfnamefont {C.}~\bibnamefont
  {Wang}}, \bibinfo {author} {\bibfnamefont {B.}~\bibnamefont {Rallabandi}}, \
  and\ \bibinfo {author} {\bibfnamefont {S.}~\bibnamefont {Hilgenfeldt}},\
  }\href {\doibase 10.1063/1.4790803} {\bibfield  {journal} {\bibinfo
  {journal} {Physics of Fluids}\ }\textbf {\bibinfo {volume} {25}},\ \bibinfo
  {pages} {022002} (\bibinfo {year} {2013})}\BibitemShut {NoStop}%
\bibitem [{\citenamefont {Rallabandi}\ \emph {et~al.}(2013)\citenamefont
  {Rallabandi}, \citenamefont {Wang},\ and\ \citenamefont
  {Hilgenfeldt}}]{Rallabandi:2013eb}%
  \BibitemOpen
  \bibfield  {author} {\bibinfo {author} {\bibfnamefont {B.}~\bibnamefont
  {Rallabandi}}, \bibinfo {author} {\bibfnamefont {C.}~\bibnamefont {Wang}}, \
  and\ \bibinfo {author} {\bibfnamefont {S.}~\bibnamefont {Hilgenfeldt}},\
  }\href {\doibase 10.1017/jfm.2013.616} {\bibfield  {journal} {\bibinfo
  {journal} {J. Fluid Mech.}\ }\textbf {\bibinfo {volume} {739}},\ \bibinfo
  {pages} {57} (\bibinfo {year} {2013})}\BibitemShut {NoStop}%
\bibitem [{\citenamefont {Rallabandi}\ \emph {et~al.}(2014)\citenamefont
  {Rallabandi}, \citenamefont {Marin}, \citenamefont {Rossi}, \citenamefont
  {K\"ahler},\ and\ \citenamefont {Hilgenfeldt}}]{rallabandi2014}%
  \BibitemOpen
  \bibfield  {author} {\bibinfo {author} {\bibfnamefont {B.}~\bibnamefont
  {Rallabandi}}, \bibinfo {author} {\bibfnamefont {A.}~\bibnamefont {Marin}},
  \bibinfo {author} {\bibfnamefont {M.}~\bibnamefont {Rossi}}, \bibinfo
  {author} {\bibfnamefont {C.~J.}\ \bibnamefont {K\"ahler}}, \ and\ \bibinfo
  {author} {\bibfnamefont {S.}~\bibnamefont {Hilgenfeldt}},\ }\href@noop {}
  {\bibfield  {journal} {\bibinfo  {journal} {Submitted to Journal of Fluid
  Mechanics}\ } (\bibinfo {year} {2014})}\BibitemShut {NoStop}%
\bibitem [{\citenamefont {Cierpka}\ \emph {et~al.}(2011)\citenamefont
  {Cierpka}, \citenamefont {Rossi}, \citenamefont {Segura},\ and\ \citenamefont
  {K\"ahler}}]{Cierpka2011}%
  \BibitemOpen
  \bibfield  {author} {\bibinfo {author} {\bibfnamefont {C.}~\bibnamefont
  {Cierpka}}, \bibinfo {author} {\bibfnamefont {M.}~\bibnamefont {Rossi}},
  \bibinfo {author} {\bibfnamefont {R.}~\bibnamefont {Segura}}, \ and\ \bibinfo
  {author} {\bibfnamefont {C.~J.}\ \bibnamefont {K\"ahler}},\ }\href {\doibase
  {10.1088/0957-0233/22/1/015401}} {\bibfield  {journal} {\bibinfo  {journal}
  {Meas. Sci. Technol.}\ }\textbf {\bibinfo {volume} {22}},\ \bibinfo {pages}
  {015401} (\bibinfo {year} {2011})}\BibitemShut {NoStop}%
\bibitem [{\citenamefont {Rossi}\ and\ \citenamefont
  {K{\"a}hler}(2014)}]{rossi2014optimization}%
  \BibitemOpen
  \bibfield  {author} {\bibinfo {author} {\bibfnamefont {M.}~\bibnamefont
  {Rossi}}\ and\ \bibinfo {author} {\bibfnamefont {C.~J.}\ \bibnamefont
  {K{\"a}hler}},\ }\href@noop {} {\bibfield  {journal} {\bibinfo  {journal}
  {Experiments in Fluids}\ }\textbf {\bibinfo {volume} {55}},\ \bibinfo {pages}
  {1} (\bibinfo {year} {2014})}\BibitemShut {NoStop}%
\bibitem [{\citenamefont {Marmottant}\ \emph
  {et~al.}(2006{\natexlab{b}})\citenamefont {Marmottant}, \citenamefont
  {Versluis}, \citenamefont {de~Jong},\ and\ \citenamefont
  {Hilgenfeldt}}]{Versluis:2006vq}%
  \BibitemOpen
  \bibfield  {author} {\bibinfo {author} {\bibfnamefont {P.}~\bibnamefont
  {Marmottant}}, \bibinfo {author} {\bibfnamefont {M.}~\bibnamefont
  {Versluis}}, \bibinfo {author} {\bibfnamefont {N.}~\bibnamefont {de~Jong}}, \
  and\ \bibinfo {author} {\bibfnamefont {S.}~\bibnamefont {Hilgenfeldt}},\
  }\href {http://www.springerlink.com/index/P208434607V34747.pdf} {\bibfield
  {journal} {\bibinfo  {journal} {Experiments in Fluids}\ }\textbf {\bibinfo
  {volume} {41}},\ \bibinfo {pages} {147} (\bibinfo {year}
  {2006}{\natexlab{b}})}\BibitemShut {NoStop}%
\bibitem [{\citenamefont {Einstein}(1926)}]{einstein1926ursache}%
  \BibitemOpen
  \bibfield  {author} {\bibinfo {author} {\bibfnamefont {A.}~\bibnamefont
  {Einstein}},\ }\href@noop {} {\bibfield  {journal} {\bibinfo  {journal}
  {Naturwissenschaften}\ }\textbf {\bibinfo {volume} {14}},\ \bibinfo {pages}
  {223} (\bibinfo {year} {1926})}\BibitemShut {NoStop}%
\bibitem [{\citenamefont {Aref}(1984)}]{aref1984stirring}%
  \BibitemOpen
  \bibfield  {author} {\bibinfo {author} {\bibfnamefont {H.}~\bibnamefont
  {Aref}},\ }\href@noop {} {\bibfield  {journal} {\bibinfo  {journal} {J. Fluid
  Mech.}\ }\textbf {\bibinfo {volume} {143}},\ \bibinfo {pages} {1} (\bibinfo
  {year} {1984})}\BibitemShut {NoStop}%
\bibitem [{\citenamefont {Ottino}(1989)}]{ottino1989kinematics}%
  \BibitemOpen
  \bibfield  {author} {\bibinfo {author} {\bibfnamefont {J.~M.}\ \bibnamefont
  {Ottino}},\ }\href@noop {} {\emph {\bibinfo {title} {The kinematics of
  mixing: stretching, chaos, and transport}}},\ Vol.~\bibinfo {volume} {3}\
  (\bibinfo  {publisher} {Cambridge University Press},\ \bibinfo {year}
  {1989})\BibitemShut {NoStop}%
\bibitem [{\citenamefont {Stroock}\ \emph {et~al.}(2002)\citenamefont
  {Stroock}, \citenamefont {Dertinger}, \citenamefont {Ajdari}, \citenamefont
  {Mezi{\'c}}, \citenamefont {Stone},\ and\ \citenamefont
  {Whitesides}}]{Stroock:2002vd}%
  \BibitemOpen
  \bibfield  {author} {\bibinfo {author} {\bibfnamefont {A.~D.}\ \bibnamefont
  {Stroock}}, \bibinfo {author} {\bibfnamefont {S.~K.}\ \bibnamefont
  {Dertinger}}, \bibinfo {author} {\bibfnamefont {A.}~\bibnamefont {Ajdari}},
  \bibinfo {author} {\bibfnamefont {I.}~\bibnamefont {Mezi{\'c}}}, \bibinfo
  {author} {\bibfnamefont {H.~A.}\ \bibnamefont {Stone}}, \ and\ \bibinfo
  {author} {\bibfnamefont {G.~M.}\ \bibnamefont {Whitesides}},\ }\href@noop {}
  {\bibfield  {journal} {\bibinfo  {journal} {Science}\ }\textbf {\bibinfo
  {volume} {295}},\ \bibinfo {pages} {647} (\bibinfo {year}
  {2002})}\BibitemShut {NoStop}%
\bibitem [{\citenamefont {Liu}\ \emph {et~al.}(2000)\citenamefont {Liu},
  \citenamefont {Stremler}, \citenamefont {Sharp}, \citenamefont {Olsen},
  \citenamefont {Santiago}, \citenamefont {Adrian}, \citenamefont {Aref},\ and\
  \citenamefont {Beebe}}]{Liu:2000iu}%
  \BibitemOpen
  \bibfield  {author} {\bibinfo {author} {\bibfnamefont {R.~H.}\ \bibnamefont
  {Liu}}, \bibinfo {author} {\bibfnamefont {M.~A.}\ \bibnamefont {Stremler}},
  \bibinfo {author} {\bibfnamefont {K.~V.}\ \bibnamefont {Sharp}}, \bibinfo
  {author} {\bibfnamefont {M.~G.}\ \bibnamefont {Olsen}}, \bibinfo {author}
  {\bibfnamefont {J.~G.}\ \bibnamefont {Santiago}}, \bibinfo {author}
  {\bibfnamefont {R.~J.}\ \bibnamefont {Adrian}}, \bibinfo {author}
  {\bibfnamefont {H.}~\bibnamefont {Aref}}, \ and\ \bibinfo {author}
  {\bibfnamefont {D.~J.}\ \bibnamefont {Beebe}},\ }\href@noop {} {\bibfield
  {journal} {\bibinfo  {journal} {J. Microelectromech. Syst.}\ }\textbf
  {\bibinfo {volume} {9}},\ \bibinfo {pages} {190} (\bibinfo {year}
  {2000})}\BibitemShut {NoStop}%
\end{thebibliography}

%

\end{document}